\documentclass[10pt,twocolumn]{article}
\usepackage{graphicx}
\usepackage{color}
\usepackage{amsmath}
\usepackage{amssymb}
\usepackage[left=4cm,right=2cm,top=2cm,bottom=2cm]{geometry}
\usepackage{bm}

\newcommand{\kp}{\mbox{$\bm{k}\!\cdot\!\bm{p}$ }}

\newcommand{\tcal}{\mathcal{T}}
\newcommand{\hc}{\mathrm{h.c.}}
\newcommand{\unit}[1]{\,\mathrm{#1}}

\begin{document}
\title{Accuracy of effective mass equation for a single and double cylindrical quantum dot} 
%\title{Conduction Band Effective Mass Equation for a Nanostructure}
\author{A. Mielnik-Pyszczorski\footnote{adam.mielnik-pyszczorski@pwr.edu.pl}, K. Gawarecki, P. Machnikowski}
\date{\small Department of Theoretical Physics,
Faculty of Fundamental Problems of Technology, \\ Wroc\l{}aw University of Science and Technology, Wybrze\.{z}e Wyspia\'{n}skiego 27, 50-370 Wroc\l{}aw, Poland}
\maketitle

\begin{abstract}
\small
In this contribution we study the accuracy of various forms of electron effective mass
equation in reproducing spectral and spin-related features of quantum dot systems. We
compare the results of the standard $8$ band \kp model to those obtained from effective
mass equations obtained by perturbative elimination procedures in various approximtions
for a cylindrical quantum dot or a system of two such dots. We calculate the splitting of
electronic shells, the electron g-factor and spin-orbit induced  spin mixing and show that
for a cylindrical 
dot the $g$-factor is reproduced very exactly, while for the two other quantities the
effective mass equation is much less accurate.
%\\PACS numbers: %85.75.-d, 03.65.Yz.
\end{abstract}

\section{Introduction}

The~$\bm{k}\cdot\bm{p}$ method in the envelope function approximation
\cite{Willatzen2009,winkler03} has been very successful in modeling carrier 
states and kinetics, influence of strain, response to external fields and other properties
of quantum dot (QD) systems \cite{Kadantsev2010,doty09,Ardelt2016b}. While the 8-band \kp
model is an established standard of nanostructure modeling, in the cases when conduction
band properties are of interest a simplified model including only the conduction band (cb)
states is often applicable \cite{Luttinger1955,Dingle1974,Maan1982}. 
In common approaches, such a single-band or two-band (if spin is included) model is
expressed by familiar Schr\"odinger- or Pauli-like equations for the envelope function,
referred to as \textit{effective mass equations}, 
leading to the popular ``particle in a box'' picture of confined carriers and, in many
cases, allowing one to reduce mesoscopic semiconductor problems to textbook exercises in
quantum mechanics. An important advantage of effective mass equations is their
considerably reduced computational cost in the cases when numerical solution is required. 

In our recent work \cite{mielnik17} we presented a systematic derivation of conduction
band effective mass equations from the 8-band \kp Hamiltonian. Various
forms of such equations emerged as subsequent approximations within the framework of
L\"owdin decoupling scheme \cite{lowdin51}. We have assessed the accuracy of these
approximations by calculating the splitting between the $s$ and $p$ states and the ground
state Zeeman splitting in two models of a lens-shaped QD with different composition profiles.
We showed that a quantitatively correct description of the lowest sector of the
electron spectrum requires a self-consistent renormalization of
the Hamiltonian parameters as well as accounting for cb non-parabolicity by
self-consistently including terms of higher order in the electron momentum. It turned out
that an accurate value of the $g$ factor is obtained only after including the 
full structure of the vb Hamiltonian in the equation, which yields 
a rather complicated equation that does not resemble the Pauli equation and does not even
allow one to separate the kinetic and Pauli terms. An equation that is rigorously derived from the 
\kp theory strictly up to order $k^{2}$ (which may correspond to the most usual notion
of an \textit{effective mass theory}) quantitatively fails in all respects.

In this contribution we continue the previous analysis in two directions. First, we extend
the study to another system geometry, a cylindrical QD, in order to gain broader evidence for our
conclusions. Second, we study the performance of the efective mass equation in reproducing
a much more sophisticated quantitative feature of the system: the spin-orbit induced
anticrossing between two nominally opposite-spin Zeeman sublevels of ground states located
in two coupled QDs. We show that the previous conclusions are essentially confirmed in the present
case of a cyindrycal QD, while for the subtle spin-orbit effect the effective mass approximation
essentially captures only the order of magnitude of the resonant splitting, with
intermediate approximations producing more accurate results than the nominally most exact
one. 

\section{Model and method}\label{model}

% \begin{figure*}
% 	\centering
% 	\includegraphics[width=.9\linewidth]{wyk-przekroj-QD-DQD-sklad-jasny}
% 	\caption{The cross-sections of the systems.}
% 	\label{fig:system}
% \end{figure*}

We consider one or two cylindrical QDs placed on a wetting
layer (WL).
% as shown in Fig.~\ref{fig:system}. 
The height of the QDs is $4.2
\unit{nm}$  and their radii are $9$ and and $11.4 \unit{nm}$ in the single- and double-QD
case, respectively. The WL thickness is $0.6 \unit{nm}$.
In the double-QD system, the distance between the QDs is $10.2 \unit{nm}$ (bottom to
bottom). The dots are displaced along the in-plane direction (i.e., off-axis) by $6
\unit{nm}$ in order to break the angular momentum conservation. The QDs and WL are
composed of pure InAs and are embedded in a pure GaAs matrix. 
The computational box is $42 \times 42 \times 27 \unit{nm}$. 

The general effective mass Hamiltonian can be written using two building elements
\cite{mielnik17}: The first one are $2\times 6$ matrices $\mathcal{T}_{i}$, $i=1,2,3$,
defined by writing 
the linear in $k$ part of the off-diagonal block of the 8-band \kp Hamiltonian
($H_{\mathrm{cv}}=H_{\mathrm{6c8v}}\oplus H_{\mathrm{6c7v}}$) in the form
$H_{\mathrm{cv}}=P\bm{\mathcal{T}}\cdot\bm{k}$ (see Ref.~\cite{mielnik17} for the explicit
definition).  The second element is the $6\times 6$ matrix (in the sense of the usual
matrix notation for the multi-band \kp theory) $\mathcal{D}$, representing the local
offset between the conduction and valence bands, keeping the full structure of the
latter. This is defined as follows. First, one writes the cb block of the 8-band
Hamiltonian, neglecting Zeeman terms, as
$H_{\mathrm{6c6c}}=\chi_{\mathrm{c}}\mathbb{I}_{2\times 2}$. Then one defines
$\mathcal{D}=\chi_{\mathrm{c}}\mathbb{I}_{6\times 6}
-H_{\mathrm{v}}$ where $H_{\mathrm{v}}$ is the $6\times 6$ valence band block of
the Hamiltonian. Both $\chi_{\mathrm{c}}$ and $H_{\mathrm{v}}$ can depend on
position but not on momenta, hence, depending on the chosen approximation, the
$k$-dependent terms in these components are either neglected or replaced by
self-consistently calculated average values in the state of interest.
Here $\mathbb{I}_{n\times n}$ is the $n\times n$ unit matrix.
Then, the effective mass Hamiltonian has the form \cite{mielnik17}
	\begin{align}
	\label{H2-T}
	\tilde{H}^{(2)} & = \sum_{jl}k_{j}P\tcal_{j}\mathcal{D}^{-1}\tcal^{\dag}_{l} P k_{l} \\
	&\quad +\frac{1}{2}\sum_{jl}\left(
	P\tcal_{j}\mathcal{D}^{-1}[k_{j},\chi_{\mathrm{c}}']\mathcal{D}^{-1}
	\tcal^{\dag}_{l}P  k_{l} +\hc\right).
	\nonumber
	\end{align} 

Eq.~\eqref{H2-T} can be taken at various levels of approximation by assuming different
approximate forms for $\mathcal{D}$. As in Ref.~\cite{mielnik17}, we consider the
following series of approximations:
(1)~\textit{bulk approximation without strain}, where $\mathcal{D}$ is diagonal and represents band
offsets in a bulk crystal; 
(2)~\textit{bulk approximation with strain}, where we additionally include band shifts due
to local strain via diagonal strain-related terms; 
(3)~\textit{semi-phenomenological approximation}, where $\mathcal{D}$ is still diagonal
and takes the constant value of the energy difference between the highest vb and lowest cb
states in the QD obtained from the 8-band \kp model (to mimic the experimental fundamental
transition energy), with an additional spin-orbit shift for the $\Gamma_{\mathrm{7v}}$ band;
(4)~\textit{off-diagonal approximation}, where all the $k$-independent elements of
$\chi_{\mathrm{c}}$ and $H_{\mathrm{v}}$ are included, thus yielding an equation that is
strictly equivalent to the 8-band model up to terms quadratic in $k$.
(5)~\textit{off-diagonal $+\left<k^2\right>$ approximation}, where the average values of the
$k$-dependent terms in $\chi_{\mathrm{c}}$ and $H_{\mathrm{v}}$ are self-consistently
included; 
(6)~\textit{self-consistent effective mass equation}, where the parameters of the original
8-band Hamiltonian are renormalized, which accounts for effects beyond perturbation theory
(see Ref.~\cite{mielnik17} for details).

\section{Results}\label{results}

\begin{figure*}[tb]
\centering
\includegraphics[width=.9\linewidth]{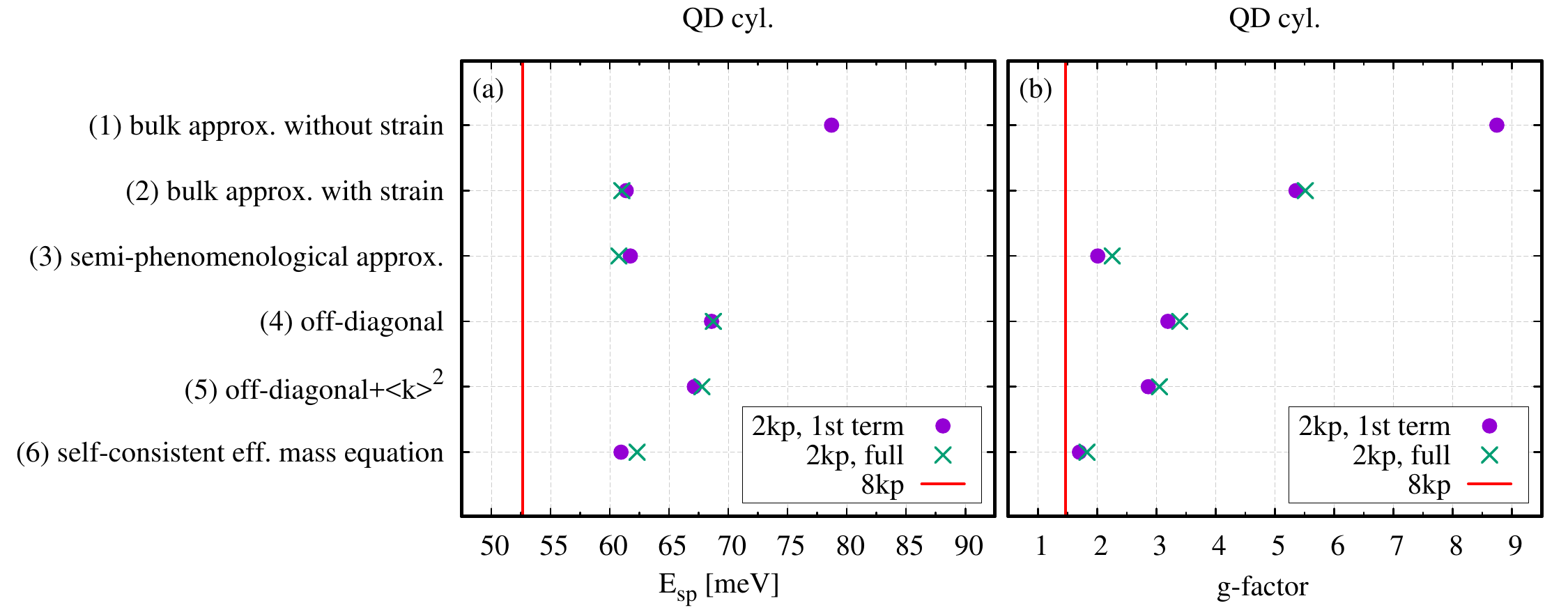}
\caption{Comparison of various effective mass approximations. (a) Energy difference
  between the ground and first excited state in a single QD at zero magnetic field. (b)
  Ground state  g-factor. In both figures the red line is the reference value from the
  8-band \kp model, dots show the results obtained by taking only the first term of
  Eq.~\eqref{H2-T} into account and crosses show the values for the full Hamiltonian in a
  given approximation.}
\label{fig:wyk-Esp_gf}
\end{figure*}
 
In Fig. ~\ref{fig:wyk-Esp_gf}(a) we present the values of the $s$-$p$ energy separation
for a single 
cyllindrical QD obtained in the approximations defined above. As in the case of 
lens-shaped QD
models discussed in Ref.~\cite{mielnik17}, the overall accuracy of the effective mass
theories is not impressive: Effective models tend to overestimate this intraband
excitation energy and even the most 
complicated approximation (6) yields the result with $\approx 15\%$ error, corresponding
to $\approx 8 \unit{meV}$ absolute difference. Unlike the models of Ref.~\cite{mielnik17},
where the most sophisticated approximation was the most accurate, here the standard
approach with simple, diagonal $\mathcal{D}$ (approximation (2) or (3)) is able to reach
the same accuracy.  

Fig.~\ref{fig:wyk-Esp_gf}(b) shows the results for the Land\'e factor of the ground state,
obtained from the Zeeman splitting at $0.1 \unit{T}$. For the most accurate approximation,
the relative error is $20\%$, which corresponds to only $0.29$ in absolute value. The
relative error is between the values obtained for two composition models in
Ref.~\cite{mielnik17}. As in the QD models studied in that previus work, the full
self-consistent model (6) yields the best results.

\begin{figure}
	\centering
	\includegraphics[width=\linewidth]{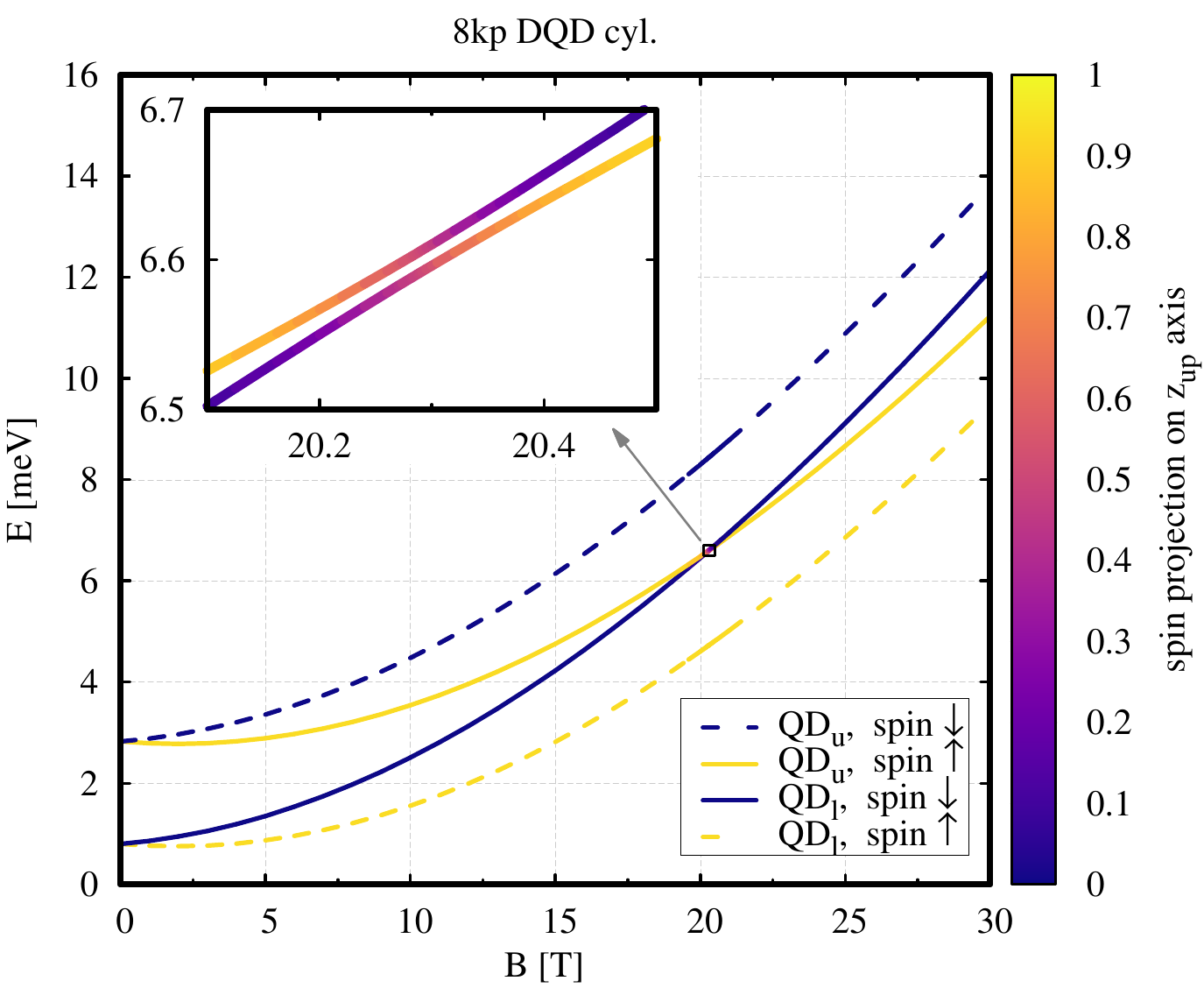}
	\caption{Dependence of electronic states on the magnetic field in the DQD structure.}
	\label{fig:wyk-DQDcyl_E-B}
\end{figure}

\begin{figure*}
	\centering
	\includegraphics[width=.6\linewidth]{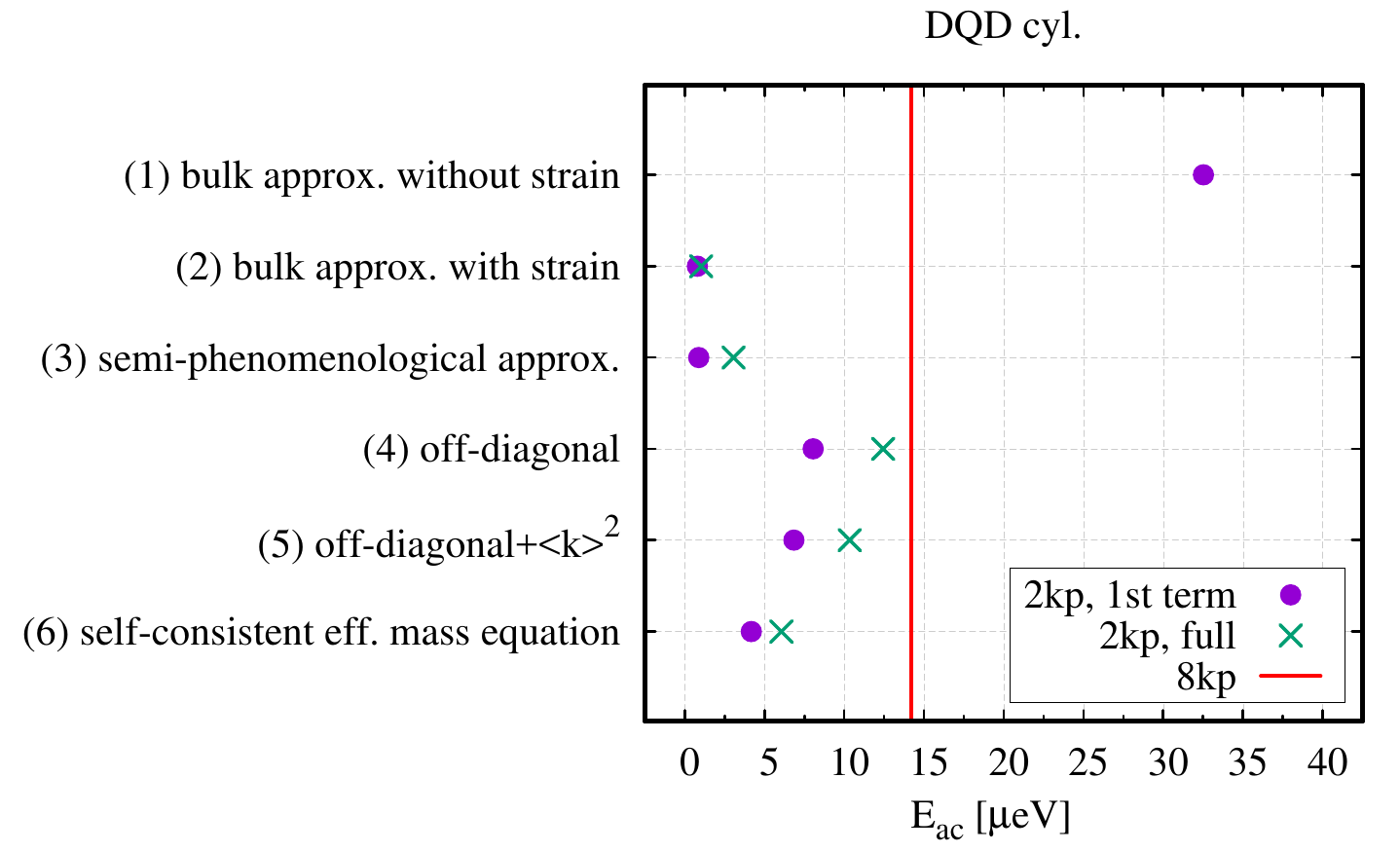}
	\caption{Comparison of the spin-orbit anticrossing width according to different
          approximations.}
	\label{fig:wyk-Eac}
\end{figure*}

Finally, we discuss spin mixing induced by spin-orbit
effects resulting from structure inhomogeneity (the usual Dresselhaus terms are switched
off in the models for the sake of clarity of the discussion). In a double-QD structure
with broken axial symmetry coupling between 
two opposite Zeeman states belonging nominally to two diferent QDs becomes possible, which
leads to an avoided crossing structure at the intersection of these two states
(Fig.~\ref{fig:wyk-DQDcyl_E-B}), thus yielding a spectral feature that allows one to
quantify the sthrength of the spin-orbit effects. The results, shown in
Fig.~\ref{fig:wyk-Eac}, indicate that even the most sophisticated equations are able to
capture at most the order of magnitude of this effect. Moreover, the model involving
self-consistent corrections, that was relatively reliable in the previous cases, produces
less accurate results than the off-diagonal model without these corrections.

\section{Conclusions}\label{conclusions}

Our results indicate that a properly constructed effective mass equation with
non-parabolicity corrections is able to quantitatively reproduce the~system
spectrum, including the~Zeeman splitting, within 30\% accuracy. On the contrary, a more
sophisticated feature related to spin-orbit coupling is not accurately reconstructed and
tend to be considerably underestimated. This
suggests in particular that effective mass models should be used with much care in the
description of admixture-induced spin relaxation processes in self-assembled structures.

\small
\bibliographystyle{prsty}
\bibliography{library}

\end{document}